\documentclass[utf8]{frontiersinFPHY_FAMS} 
\setcitestyle{square} 
\usepackage{url,hyperref,lineno,microtype}
\usepackage[onehalfspacing]{setspace}
\usepackage{amscd,amsmath,amsthm,amssymb,color}
\usepackage[arrow,matrix]{xy}
\usepackage{graphicx}
\usepackage{epstopdf}
\usepackage{soul}
\usepackage{xcolor}

\newcommand{\bea}{\begin{eqnarray}}
\newcommand{\eea}{\end{eqnarray}}
\newcommand{\bes}{\begin{subequations}}
\newcommand{\ees}{\end{subequations}}

\newcommand{\sech}{\mbox{sech}}

\def\firstAuthorLast{K. Sakkaravarthi {et~al.}} 
\def\Authors{Karuppaiya Sakkaravarthi\,$^{1}$, Sudhir Singh\,$^{2}$ and Natanael Karjanto\,$^{3,*}$}

\def\Address{$^{1}$Young Scientist Training Program, Asia-Pacific Center for Theoretical Physics (APCTP), POSTECH Campus, Pohang - 37673, South (Republic of) Korea \\
$^{2}$Department of Mathematics, National Institute of Technology, Tiruchirappalli - 620015, Tamil Nadu, India\\
$^{3}$Department of Mathematics, University College, Natural Science Campus Sungkyunkwan University, Suwon - 16419, South (Republic of) Korea}
\def\corrAuthor{Natanael Karjanto} 
\def\corrEmail{Email: natanael@skku.edu}

\begin{document}
\onecolumn
\firstpage{1}
	
\title[Exploring the Dynamics of Nonlocal Nonlinear Waves]{Exploring the Dynamics of Nonlocal Nonlinear Waves: Analytical Insights into the Extended Kadomtsev-Petviashvili Model}

\author[\firstAuthorLast ]{\Authors} 
\address{} 
\correspondance{} 
	
\extraAuth{}
	
\maketitle 
\setstretch{1.250}
\begin{abstract}
The study of nonlocal nonlinear systems and their dynamics is a rapidly increasing field of research. In this study, we take a closer look at the extended nonlocal Kadomtsev-Petviashvili (enKP) model through a systematic analysis of explicit solutions. Using a superposed bilinearization approach, we obtained a bilinear form of the enKP equation and constructed soliton solutions. Our findings show that the nature of the resulting nonlinear waves, including the amplitude, width, localization, and velocity, can be controlled by arbitrary solution parameters. The solutions exhibited both symmetric and asymmetric characteristics, including localized bell-type bright solitons, superposed kink-bell-type and antikink-bell-type soliton profiles. The solitons arising in this nonlocal model only undergo elastic interactions while maintaining their initial identities and shifting phases. Additionally, we demonstrated the possibility of generating bound-soliton molecules and breathers with appropriately chosen soliton parameters. The results of this study offer valuable insights into the dynamics of localized nonlinear waves in higher-dimensional nonlocal nonlinear models.
{\section{Keywords:} Nonlocal Nonlinear Model; (2+1)D Extended Kadomtsev-Petviashvili Equation; Nonlinear Wave Solutions; Soliton Interaction; Bound States; Breathers.} 
\end{abstract}
	
\section{Introduction}
	
The study of nonlinear waves has been a captivating subject for many researchers for centuries. With recent advancements in mathematical methods, computational techniques, and experimental demonstrations, the investigation of nonlinear waves has gained increasing attention~\cite{Yang-book}. Described by nonlinear partial differential equations (PDEs), the study of nonlinear waves encompasses a wide range of disciplines, including mathematics, physics and engineering. As computational power and data analysis methods continue to improve, researchers are exploring new techniques to leverage the potential of nonlinear waves in various fields such as telecommunications, optics, and material sciences~\cite{Yang-book,infeld2001nonlinear, ablowitz2011nonlinear}. The exploration of these waves requires a combination of precise mathematical modeling, solving equations, analyzing their dynamics, and verifying the results through experiments.
	
Solitons are unique self-sustaining waves that maintain their shape and velocity as they propagate. Unlike other waves that tend to spread out or change form, solitons maintain their characteristics even as they move through various systems, including water, plasma, and optical fibers. Mathematically, solitons are solutions to nonlinear PDEs in which the nonlinearities of the underlying physical systems balance the dispersive effects. These solutions exhibit notable properties such as energy conservation and unchanging waveforms~\cite{ablowitz1981solitons}. Although first discovered in water waves, solitons have since been observed in many other areas of study, including plasma physics, nonlinear optics, and even some biological systems. However, solitons are just one type of nonlinear waves with distinct properties, and exploring their dynamics holds tremendous promise for a multitude of applications, including telecommunications, optics, Bose-Einstein condensates, water waves, {and biological systems}~\cite{Yang-book,infeld2001nonlinear, ablowitz2011nonlinear,ablowitz1981solitons,eilenberger1983solitons,kavallaris2018non}. 
	
The study of nonlinear waves is a fascinating field that encompasses various models, including the Korteweg-de Vries (KdV), Kadomtsev-Petviashvili (KP), sine-Gordon (sG), Boussinesq, and nonlinear Schr\"odinger (NLS) equations~\cite{Yang-book,infeld2001nonlinear, ablowitz2011nonlinear,ablowitz1981solitons}. The KP equation, introduced by Boris Borisovich Kadomtsev~(1928--1998) and Vladimir Iosifovich Petviashvili~(1936--1993), models the propagation of shallow-water waves in the presence of weakly nonlinear restoring forces and frequency dispersions \cite{kadomtsev1970stability}. This is a two-dimensional integrable generalization of the classical KdV equation, which describes the propagation of unidirectional shallow-water waves. Both KP and KdV equations are examples of completely integrable systems and admit a range of solutions, including $N$-soliton solutions, Lax pairs, solvable inverse scattering transform (IST), Painlev\'e integrability, and exact analytical solutions for localized, interacting, periodic, elliptic, and quasi-periodic nonlinear waves. Additionally, they have an infinite number of conservation laws~\cite{konopelchenko1993solitons,johnson1997modern,lou1997infinitely}.
	
The KdV equation, introduced by Dutch mathematicians Diederik Johannes Korteweg~(1848--1941) and Gustav de Vries~(1866--1934) in 1895, is a mathematical model that effectively describes the long-term evolution of dispersive wave phenomena. In the KdV equation, the nonlinear influence of wave steepening is balanced by the dispersion effect~\cite{debnath2012nonlinear}. The KP equation is an extension of the KdV equation and includes wave dynamics in the transverse direction. The KdV and KP equations can be written as follows respectively \cite{Yang-book,infeld2001nonlinear,ablowitz2011nonlinear,ablowitz1981solitons,eilenberger1983solitons,kavallaris2018non,kadomtsev1970stability,konopelchenko1993solitons,johnson1997modern,lou1997infinitely,debnath2012nonlinear}: 
\begin{align}
u_t - 6 u u_{xx} + u_{xxx} &= 0, \\ 
\left(u_t - 6 u u_{xx} + u_{xxx} \right)_x + 3 \sigma^2 u_{yy} &= 0.
\end{align}

The wave functions $u = u(x,t)$ and $u = u(x,y,t)$ are both scalar fields, where $x$, $y$, and $t \in \mathbb{R}$ represent the longitudinal spatial coordinate, transverse spatial coordinate, and temporal variable, respectively. The subscripts indicate partial derivatives with respect to the associated variables. The KP equation is commonly used to model water waves with weak surface tension (KP~II equation, with $\sigma = 1$) or waves in thin films with strong surface tension (KP~I equation, with $\sigma = i$)~\cite{ablowitz2011nonlinear,biondini2008KP,horikis2017light}. In addition to water waves~\cite{ablowitz1979evolution,segur1985analytical,hammack1989two,hammack1995two}, the KP equation has been applied as a model for other physical systems, such as nonlinear optics~\cite{pelinovsky1995self}, ferromagnetic media~\cite{leblond2002kp}, plasma physics~\cite{alharbi2020analytical}, Bose-Einstein condensates~\cite{tsuchiya2008solitons}, and many more. 

Ma et al. (2021) developed an extended yet integrable KP equation and obtained soliton, breather, and lump interaction solutions using the Hirota bilinear form~\cite{ma2021new}. In a separate study, they extended the integrable KP equation to a 3D case and observed abundant dynamic behaviors, such as fusing and splitting phenomena in lump wave interactions~\cite{ma2021anew}. {Recently, Rao et al. (2021) have studied resonant collisions among localized lumps and line solitons of the KPI equation, resulting in the discovery of ``rogue lumps'', which are completely localized in both space and time and possess the features of two-dimensional rogue waves. Analytically, these solutions are constructed using the KP hierarchy reduction method and Hirota's bilinearization procedure, and the phenomena of resonant interactions and rogue lumps are expected to persist for other nonlinear evolution equations with two spatial dimensions~\cite{rao2021completely}.}
 
Whereas solitons in one-dimensional systems have already garnered significant attention from researchers due to their intriguing complexity, solitons in higher-dimensional settings often display more complex dynamics arising from the interplay between different types of nonlinearity, dispersive effects, geometric factors, and varying nonlocalities. Carr and Brand (2008) explored the extension of 1D solitons into 2D and 3D in a book chapter, highlighting that trapped BECs can stabilize these solitons and lead to the emergence of new nonlinear objects and topological solitons, depending on whether the nonlinearity is repulsive or attractive~\cite{carr2008multidimensional}. Mihalache (2021) provided a comprehensive overview of recent theoretical and experimental studies on localized structures in optical and matter-wave media, covering various physical contexts such as light bullets, solitons, and rogue waves~\cite{mihalache2021localized}. 
Malomed's (2016) review on multidimensional solitons discussed the stabilization of soliton states in both 2D and 3D physical settings. In 2D, the solitons form a stable ground state, whereas in 3D, they form metastable solitons~\cite{malomed2016multidimensional}. In his latest monograph, published in 2022, Malomed addressed the potential benefits of extending solitons from 1D to higher-dimensional systems while highlighting the challenges, particularly regarding stability. These challenges open up many interesting and unresolved problems in the area of nonlinear waves~\cite{malomed2022multidimensional}.
	
The NLS equation is a well-known integrable system that was initially proposed as a nonlocal nonlinear evolution equation. This equation has an infinite number of conservation laws in the context of Hamiltonian dynamical systems, making it integrable and also exhibiting parity-time (${\mathcal {PT}}$) reversal symmetry in its self-induced potential~\cite{ablowitz2013integrable,ablowitz2016inverse}. The feature of nonlocality in the NLS equation can appear in the spatial, temporal, or both spatial and temporal variables. Several other integrable nonlocal nonlinear evolution equations have been identified, which can be obtained through simple symmetry reductions of the general Ablowitz--Kaup--Newell--Segur~(AKNS) hierarchy that is solvable by the inverse scattering transformation~\cite{ablowitz1974inverse}. In $(1 + 1)$-dimensions, other nonlinear evolution models such as the derivative NLS, modified Korteweg-de Vries (mKdV), and sine-Gordon equations also exhibit nonlocality. In $(2 + 1)$-dimensions, other PDE models such as the three-wave interaction, generalized NLS, and Davey–Stewartson (DS) equations display this nonlocallity property too~\cite{ablowitz2017integrable}.

Research on nonlocal models in the NLS equation and its related models, such as the Gross-Pitaevskii (GP) equation, has gained momentum in the third decade of this century. For example, Li et al. (2020) presented one- and two-bright-soliton solutions of the generalized nonlocal GP equation with an arbitrary time-dependent linear potential obtained through the improved Hirota method, which can describe the dynamics of soliton solutions in quasi-one-dimensional BEC and may be useful for understanding physical phenomena in nonlinear nonlocal soliton equations~\cite{li2020a}. Interestingly, Yu et al. (2021) combined the feature of both locality and nonlocality in the NLS equation and obtained its one- and two-soliton solutions using Hirota’s bilinearization transformation. This mixed local--nonlocal nonlinear system can be useful for understanding physical phenomena in nonlinear wave propagation and applied in fields such as nonlinear optics and meteorology~\cite{yu2021broken}. Recently, Li et al. (2023) presented a novel application of Darboux transformation for solving the $(2+1)$-dimensional nonlocal NLS equation with reverse time field $q(x,y,-t)$, deriving various soliton solutions on a background of kink waves, which can be extended to other nonlocal nonlinear or higher-dimensional soliton equations~\cite{li2023some}.

Apart from the above NLS type nonlocal systems, the KP equation has been the subject of investigation for nonlocal models and their corresponding dynamics, with rich variations of wave motion observed in lump solutions, line breathers, and periodic normal breathers~\cite{zhang2017breather}. These nonlocal KP equations were inspired by Lou's Alice--Bob (AB) system, which explored two-place physics and multi-place physics~\cite{lou2016alice,lou2017alice,lou2018alice}. Lou and Huang's pioneering work in 2017 established that an extension to shifted parity and delayed time reversal was advantageous for seeking group-invariant solutions for such models~\cite{lou2017alice}. 
	
Recent studies in the published literature have employed the Hirota bilinear technique to explore various extended KP equations and nonlocal AB systems, highlighting the integrability of these systems. Despite the variety of solutions obtained---including solitons, breathers, lumps, and rogue waves---they share a common approach. Among the novel solutions discovered, some exhibit intriguing dynamical behaviors, such as semi-rational solutions with elastic interactions and symmetry-breaking solitons. For instance, Manukure et al. (2018) provided conditions for the rational localization of solutions and presented lump solutions for an extended KP equation~\cite{manukure2018lump}. Fei et al. (2019) derived the nonlocal AB-Boussinesq system for surface gravity waves, investigated its residual and finite symmetries, and obtained several exact solutions~\cite{fei2019controllable}. Wu and Lou (2019) also found solitons, breathers, and lump solutions using shifted parity and delayed time reversal to explore the exact solutions of the AB-KP equation~\cite{wu2019exact}.
	
Guo et al. conducted two studies using the Hirota bilinear technique to investigate an extended KP equation. In their 2020 study, they obtained exact analytical expressions for the higher-order soliton and breather solutions of the equation and discovered novel semi-rational solutions that exhibit elastic interactions~\cite{guo2020exact}. In their 2021 study, they constructed multiple-order line rogue wave solutions using the Hirota bilinear method and a symbolic computation approach~\cite{guo2021multiple}. 

Shen et al. (2020) investigated the nonlocal AB Benjamin-Ono system using parity and time-reversal symmetry reduction and obtained breather, lump, and symmetry-breaking soliton solutions via an extended B\"acklund transformation and Hirota bilinear form~\cite{shen2020abundant}. Cao et al. (2021) studied another nonlocal ABKP system and proposed a constant dependence in the B\"acklund transformation to solve the system using the Hirota bilinear form. They found that this constant affects the symmetry-breaking characteristics of the solutions and obtained various analytical solutions, including single solitons, breathers, lumps, entangled lumps, and a pair of stripe solutions~\cite{cao2021symmetry}. Recently, Dong et al. (2023) presented an integrable AB system for the $(1 + 1)$-dimensional Boussinesq equation with shifted parity and delayed time-reversal symmetry. To obtain explicit solutions, including line solitons, breathers, and lumps, the authors introduced an extended Bäcklund transformation~\cite{dong2023the}.
	
In this study, we aim to investigate the nonlocal version of the $(2 + 1)$-dimensional extended Kadomtsev-Petviashivili (eKP) equation introduced by Manukure et al. (2018), which they solved to obtain the lump wave solution using the Hirota bilinear method with quadratic polynomials~\cite{manukure2018lump}. The eKP equation is given by
\bea 
\left(u_t+6 u u_x+u_{x x x}\right)_x-u_{y y}+\alpha u_{t t}+\beta u_{y t}=0,    \label{eKPeq}
\eea 
where $x$, $y$, and $t$ represent the two spatial dimensions and time variables, respectively. It features second-order temporal and spatio-temporal dispersion coefficients $\alpha$ and $\beta$, respectively. While the equation is non-integrable for arbitrary $\alpha$ and $\beta$ values, it becomes Painlev\'e integrable when $\alpha=-\beta^2/4$~\cite{ma2021new}. The literature reports various exotic nonlinear wave solutions for this model, such as solitons and breathers~\cite{guo2020exact}, multi-line rogue waves~\cite{guo2021multiple}, and interacting wave structures~\cite{ma2021new}. 

Although the local eKP equation (\ref{eKPeq}) has been extensively studied, there has been little investigation of its nonlocal counterpart, which is the focus of our research. We seek to utilize the AB approach to construct exotic nonlinear wave solutions, such as solitons and lumps, for the nonlocal version of the eKP equation~\eqref{eKPeq} and examine their evolutionary dynamics under nonlocal effects to uncover physically interesting wave phenomena. Our objective is to contribute to the understanding of the nonlocal eKP equation and provide insights into its potential applications in various fields.
	
The remainder of this article is organized as follows. In Section~\ref{sec-Back}, we derive the nonlocal version of the eKP equation~\eqref{eKPeq} and provide its B\"acklund transformation and bilinear formalism. Section~ \ref{sec-Sol} presents one- and two-soliton solutions, including their propagation characteristics and symmetric and asymmetric interactions, as well as bound states and breathers. In Section~ \ref{sec-Lump}, we introduce a rational and well-localized lump wave solution. Finally, we conclude with a summary of our findings in the final section.
 
\section{B\"acklund Transformation and Bilinear form}   \label{sec-Back}
	
By implementing the AB physics approach, the nonlocal counterpart of the considered equation (\ref{eKPeq}) can be obtained through the substitution $u(x,y,t)=\frac{1}{2}\left(P(x,y,t)+Q(x,y,t)\right)$:
\begin{align}
\alpha (P_{tt} + Q_{tt} ) & + \beta (P_{yt} + Q_{yt} ) + (P_{xt} + Q_{xt}) - (P_{yy} + Q_{yy} ) \notag \\
&  + 3 (P_x + Q_x ) ^2 + 3 (P+Q) (P_{xx} + Q_{xx} ) + (P_{xxxx} + Q_{xxxx} ) =0. \label{eq1} 
\end{align}

Further, it can be explicitly written in the following two-coupled equations: 
\begin{subequations}
\begin{align}
\alpha P_{tt} + \beta P _{yt} + P_{xt} - P_{yy} + P_{xxxx} + \dfrac{3}{2} (P_x+Q_x)^2 + 3 P (P_{xx} + Q_{xx} )   + H(P,Q) & = 0, \label{eq2a}\\
\alpha Q_{tt} +  \beta Q_{yt} + Q_{xt} - Q_{yy} + Q_{xxxx} +  \dfrac{3}{2} (P_x+Q_x)^2 + 3 Q(P_{xx} + Q_{xx} ) - H(P,Q)&=0, \label{eq2b}
\end{align} 
\end{subequations}
where $Q$ is related to $P$ through $Q=P_s^x P_s^y P_d^t=P(-x+p_0, -y+q_0, -t+r_0)$, $P_s T_d$ are parity transformation and time reversal, respectively, and $H(P,Q)$ is an arbitrary function of both $P$ and $Q$, which can have infinitely many choices. Moreover, we choose the following form for $H(P,Q)$:
\begin{equation}
H(P,Q) = \dfrac{3}{4} (P_x^2 - Q_x^2  - (P-3Q)P_{xx} + (Q-3P)Q_{xx} ). \label{eq3}
\end{equation}

Finally, we obtain the required Alice-Bob nonlocal version of the eKP equation (\ref{abekp}), which we refer to as the extended nonlocal KP (enKP) model, given as follows:
\bes\bea 
4 (\alpha P_{tt} + \beta P_{yt} + P_{xt} - P_{yy}+ P_{xxxx }) + 3(P_x +Q_x)(3 P_x+ Q_x) + 3(P+Q)(3P_{xx} + Q_{xx}) =0,  \label{eq4}\\
4 (\alpha Q_{tt} + \beta Q_{yt} + Q_{xt} - Q_{yy}+ Q_{xxxx }) + 3(P_x +Q_x)(P_x+3 Q_x) +3(P+Q)(P_{xx} +3 Q_{xx})  =0. \label{eq5}
\eea \label{abekp}\ees 
It is interesting to check the integrability of the model, for which we have performed the Painlev\'e analysis \cite{steeb} for the above equation (\ref{abekp}) in the local setting for simplicity.

From the result, we found that the above coupled equation  (\ref{abekp}) becomes non-integrable not only for arbitrary $\alpha$ and $\beta$ parameters (as expected) but also for $\beta=-\alpha^2/4$, for which the single-component eKP equation (\ref{eKPeq}) is integrable. We do not present the details of the Painlev\'e analysis for the sake of space and its not-so-interesting mathematical equations. 

One can also view the above enKP model (\ref{abekp}) as a particular form of coupled KP systems. For example, the enKP system (\ref{abekp}) can be obtained as a special case from the following generalized version of coupled KP type equation (similar to the equations studied in Refs. \cite{lou2018alice, wuhy}) for an appropriate choice of $\gamma_j, j=0,1...10,$ coefficients:
\bes\bea 
\gamma_0 P_{tt}+[(\gamma_1 P+\gamma_2 Q)_t+(\gamma_3 P+\gamma_4 Q)_{xxx}+(\gamma_5 P+\gamma_6 Q)P_x \qquad \qquad \notag  \\+(\gamma_7 P+\gamma_8 Q)Q_x]_x +\gamma_9 P_{yy}+\gamma_{10} P_{yt} =0,  \label{eq400}\\
\gamma_0 Q_{tt}+[(\gamma_1 Q+\gamma_2 P)_t+(\gamma_3 Q+\gamma_4 P)_{xxx}+(\gamma_5 Q+\gamma_6 P)Q_x \qquad \qquad \notag  \\+(\gamma_7 Q+\gamma_8 P)P_x]_x +\gamma_9 Q_{yy}+\gamma_{10} Q_{yt} =0.\label{eq50}
\eea \label{eq50a}\ees 
To be precise, our enKP model (\ref{abekp}) can be reduced from the above equation (\ref{eq50a}) for the choice $\gamma_0=4\alpha$, $\gamma_1=\gamma_3=4$, $\gamma_2=\gamma_4=0$, $\gamma_5=\gamma_6=9$, $\gamma_7=\gamma_8=3$, $\gamma_9=-4$, and $\gamma_{10}=4\beta$. In order to solve our intended enKP model (\ref{abekp}), we adopt Hirota's bilinearization method \cite{Hirota-book, ksjpa11, kspre14, ksps20, kscsf23}, which is one of the efficient analytical techniques available for solving different classes of nonlinear equations. For this purpose, we consider the following superposed variable transformation:
\bes
\bea
&& P = 2 (\ln F )_{xx} + A (\ln F)_x, \\
&& Q = 2 (\ln F  )_{xx} - A (\ln F ) _{x},
\eea \label{eq8}
\ees
\!\!\!
where $F(x,y,t)$ is the required function to be determined while $A$ is an arbitrary real constant. It is clear to find that one can obtain the standard bilinearizing transformation when $A=0$. However, we utilize the superposed bilinear transformation with $A\neq 0$ to deduce the required bilinear equation(s). 

Note that the superposed bilinearization (\ref{eq8}) is quite different from the standard bilinearization procedure \cite{Hirota-book, ksjpa11, kspre14, ksps20, kscsf23}, and there are a few recent reports on similar superposed techniques in the literature~\cite{lou2016alice,lou2017alice,lou2018alice,cao2021symmetry}. 

On applying the above bilinearizing transformation to Eq. (\ref{abekp}) and after its simplification, we get the following bilinear equation in a standard $D$-operator form: 
\begin{equation}
(D_x D_t + D_x ^4 - D_y^2 + \alpha D_t ^2  + \beta D_y D_t) F \cdot F = 0,\label{eq9}
\end{equation}
where $D_j,~j=x,y,t,$ are standard Hirota derivatives \cite{Hirota-book}. 

Further, the above bilinear equation can be written in its equivalent expanded form as below for a clear inference of Hirota derivatives. 
\begin{align}
(FF_{xt} - F_x F_t) + (FF_{xxxx} - 4 F_x F_{xxx} + 3 F_{xx}^2) - (FF_{yy} - F_y^2 ) \qquad \nonumber\\ + \alpha (FF_{tt} - F_t^2 )+ \beta (FF_{yt}- F_yF_t)  =0. \label{eq10} 
\end{align}
Thus from the above bilinear equation (\ref{eq9}) or its equivalent form (\ref{eq10}), one can find a number of exact solutions to the considered enKP Eq. (\ref{abekp}) through different forms of explicit $F(x,y,t)$ computed for appropriate initial seed solutions.
\setstretch{1.450}
 
\section{Dynamics of Solitons and their Interactions}\label{sec-Sol}
	
In this section, we construct exact soliton solutions of the considered enKP model (\ref{abekp}) and explore their propagation as well as interaction dynamics in detail for different choices of parameters. 
	
\subsection{First-order Solitons}
To obtain the first-order (or one) soliton solution of the considered model, we consider the following form of initial seed solution: 
\begin{equation}
F(x,y,t) = \cosh (\eta _1(x,y,t)), \label{eq11}
\end{equation}
where $\eta _1(x,y,t) = k_1 \left (\left ( x - {p_0}/{2} \right ) + l_1 \left ( y - {q_0}/{2} \right ) + m_1 \left ( t - {r_0}/{2} \right ) \right )$. On applying the seed (\ref{eq11}) into the bilinear form (\ref{eq8}) and solving it, we obtain the explicit form of $m_1$ as $m_1= \left({-1 - \beta l_1 \pm  \sqrt{(1+ \beta l_1 )^2 - 4 \alpha ( 4 k_1 ^2 -l_1 ^2 ) }}\right)/{2 \alpha} $. Thus, from the dependent variable transformation (\ref{eq10}), we obtain the exact form of the first-order soliton solution as given below.
\begin{subequations}
\begin{align}
P(x,y,t) & = 2 k_1 ^2 \sech ^2 (\eta _1 ) + A k_1 \tanh (\eta _1 ), \label{eq12a}\\
Q(x,y,t) & = 2 k_1 ^2 \sech ^2 (\eta _1 ) - A k_1 \tanh (\eta _1). \label{eq12b}
\end{align} \label{eq12}
\end{subequations}
	
The above solution preserves symmetry under both spatial ($x-y$) inversion-shift and time reversal, which can be represented as $Q(x,y,t)=P(-x+p_0, -y+q_0, -t+r_0)$, and quantifies that the dynamics of mode $Q$ can be explored straightforwardly through the space-time reversal operation on the mode $P$. 

The solution (\ref{eq12}) has five arbitrary parameters ($k_1$, $l_1$, $A$, $\alpha$, and $\beta$) in addition to space-time inversion parameters $p_0$, $q_0$, and $r_0$. Here the first two parameters $k_1$ and $l_1$ represent the amplitude of the bell-kink soliton and spatial localization (soliton angle), respectively. However, among the latter, the parameter $A$ denotes the presence and absence of superposed kink-wave with bell-type soliton, and its magnitude contributes to the amplitude of the background wave. The parameters $\alpha$ and $\beta$ are real system quantities arising from the model (\ref{abekp}), and they control the second-order temporal and spatio-temporal dispersion, which shift the resulting soliton in spatial and time domain. It is essential to note from the above solution (\ref{eq12}) that the superposed bilinear transformation enables us to obtain rich characteristics arising from both localized bell (sech) and kink (tanh) type wave structures. Here the amplitude of bright bell-type soliton is described by $2k_1^2$, while $A k_1$ defines that of the background kink-soliton. At the same time, their velocity along $xt$- and $yt$-planes are defined by $-m_1$ and $-m_1/l_1$, respectively, where $m_1$ takes the form as given below Eq. (\ref{eq11})  involving all arbitrary parameters. Further, the amplitude defining parameter $k_1$ along with $l_1$, $\alpha$ and $\beta$ influence the position of the soliton at any given time. 
\begin{figure}[ht]
\centering
\includegraphics[width=0.85\linewidth]{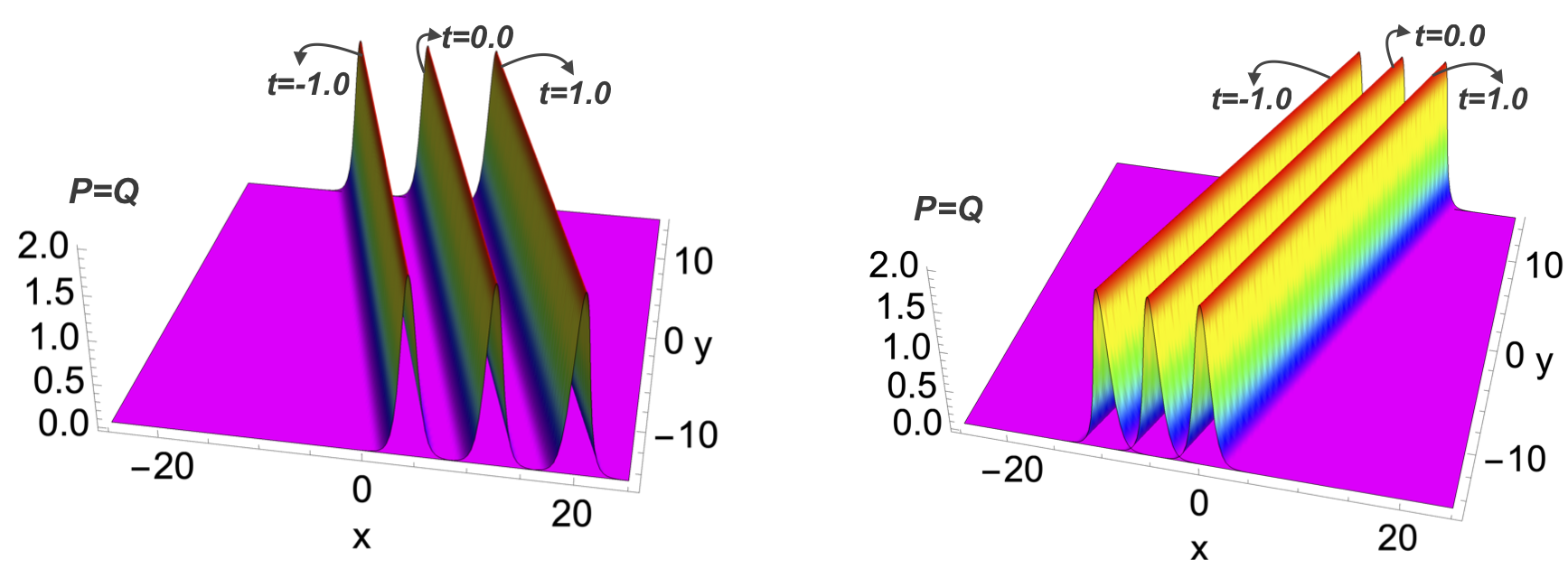}
\caption{The evolution of symmetric bell-type bright soliton ($P=Q$) arising for vanishing background ($A=0$, absence of superposed wave) at three different times ($t=-1.0,~0,~1.0$) for $l_1=0.5$ (left panel) and $l_1=-0.5$ (right panel) with other values are kept same as $\alpha=-0.25$, $\beta=1$, $k_1=1$, and $p_0=q_0=r_0=1$.}
\label{fig-1soli-1}
\end{figure} 
	
\begin{figure}[ht]
\centering
\includegraphics[width=0.85\linewidth]{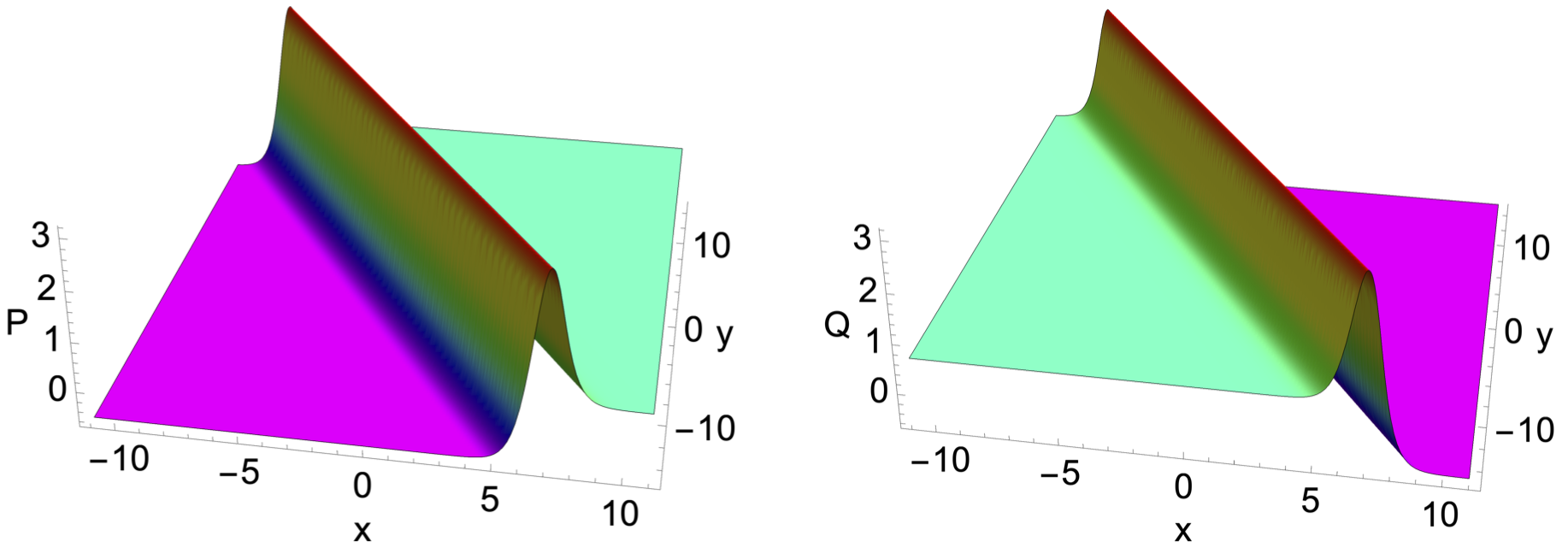}
\caption{Nature of asymmetric structured one-soliton evolution resulting in the combined bright bell-type and kink-type soliton in mode $P$ for $A=0.5$ with $\alpha=-0.5$, $\beta=1$, $k_1=1.25$, $l_1=0.5$ and $p_0=q_0=r_0=1$ at $t=0.75$. The evolution in $Q$ is nothing but the nonlocal symmetry-preserving pattern of the mode $P$ revealing the superposed bright soliton with anti-kink soliton.} \label{fig-1soli-2}
\end{figure} 
	
One of the exciting features of the present solution (\ref{eq12}) depends on the superposition parameter $A$ by admitting symmetric or asymmetric solitons based on its absence ($A=0$) or its presence ($A\neq 0$), respectively. Here we classify the soliton as symmetric and asymmetric purely based on the nature (appearance) of the profile structure. We found that the symmetric soliton results in a bell-type soliton (the so-called bright soliton, a localized and stable profile on zero background). We have shown such symmetric profiled bright soliton evolution in Fig. \ref{fig-1soli-1} for the parametric choice $\alpha=-0.25$, $\beta=1$, $k_1=1$, and $l_1=0.5$ ($l_1=-0.5$) by keeping the nonlocal parameters $p_0=q_0=r_0=1$. Further, this symmetric property of the present eKP system can be interpreted as degenerate-type solitons when their nature and amplitude are precisely the same in both modes $P$ and $Q$ (i.e., $P=Q$) at any specific time. 

However, in the latter case, asymmetric soliton resulting from the non-vanishing background parameter ($A\neq 0$) leads to the formation of a superposed profile comprising both bell and kink solitons. This superposed wave structure can be referred to as bell-type bright soliton appearing on a kink-soliton and anti-kink soliton backgrounds, respectively, in components $P$ and $Q$ or vice-versa. Contrary to the previous case, the present asymmetric soliton leads to non-degenerate structures with distinct natures in each component. For a better understanding, we have depicted an asymmetric superposed soliton in Fig. \ref{fig-1soli-2} for the choice $\alpha=-0.5$, $\beta=1$, $k_1=1.25$, and $l_1=0.5$ with $A=0.5$. Note that superposed solitons in both components become equal under the nonlocal (two space-reversal and time-inversion) symmetry transformation $Q=P_s^x P_s^y P_d^t=P(-x+p_0, -y+q_0, -t+r_0)$, which can also be visualized from the demonstrated Figure \ref{fig-1soli-2}. 

We wish to note despite the fact that the two-coupled enKP model (\ref{abekp}) is non-integrable (at least in the Painlev\'e sense, as mentioned in Section 2), it is still possible to observe soliton-like behavior in the obtained solutions to the equation (\ref{abekp}). These ``solitons" are not true solitons in the strict mathematical sense, but rather they are solitary waves that behave in a similar manner to solitons, like the elastic nature of collision (as shown in the forthcoming Section 3.2.1) where they reappear without changing their shape or amplitude and emerge from the collision essentially unchanged. For the sake of simplicity and ease of understanding, it is often convenient to refer to these waves as solitons, even though they are not strictly solitons.

\subsection{Second-order Soliton Solution and Interaction Dynamics}
	
Next, we construct a two-soliton solution to the enKP model (\ref{abekp}). For this purpose, we extend the similar procedure as performed in the case of first-order soliton solution by considering the following form of initial seed solution for $F$: 
\bes
\begin{equation}
F(x,y,t) = B_1 \cosh (\eta _1 + \eta _2 ) + B_2  \cosh (\eta _1 - \eta _2 ), \label{eq13}
\end{equation}
where $\eta _j(x,y,t) = k_j \left (\left ( x - {p_0}/{2} \right ) + l_j \left ( y - {q_0}/{2} \right ) + m_j \left ( t - {r_0}/{2} \right ) \right ), ~j=1,2$, while $B_1$ and $B_2$ are functions to be determined. On substituting the above seed into the bilinear equation (\ref{eq9}) or (\ref{eq10}) and solving categorically, we obtain the explicit form of amplitude parameters as below,
\bea 
B_1 & =& \sqrt{(1+ \beta l_1 )(1+\beta l_2 ) + 4 \alpha (12 k_1 k_2 -8 (k_1^2 +k_2^2) + l_1 l_2 ) -\gamma_1 \gamma_2  }, \\
B_2 & =& \sqrt{(1+\beta l_1 ) ( 1+\beta l_2 ) -4 \alpha ( 12 k_1k_2 + 8 (k_1^2 + k_2^2 )  -  l_2 l_2 ) - \gamma_1 \gamma_2}, \label{eq14}
\eea 
where the quantities $\gamma_1$ and $\gamma_2$ are identified as
\bea
\gamma _j & =& \sqrt{ 1 + 4 \alpha (l_j^2-4k_j^2) + 2\beta l_j+\beta^2 l_j^2  }, \quad j=1,2.
\eea  
In the above solution, the form of $m_j,~j=1,2,$ reads as below.
\bea 
m_j= \left({-1 - \beta l_j +  \sqrt{(1+ \beta l_j )^2 - 4 \alpha ( 4 k_j ^2 -l_j ^2 ) }}\right)/{2 \alpha}, \qquad j=1,2.
\eea \label{twosol}
\ees 
Thus, upon substituting the above explicit expression for $F$ into the dependent variable transformation (\ref{eq8}), the explicit form of the required two-soliton solution can be obtained as shown below: 
\bes \bea
P(x,y,t)&=&2\left[\frac{(k_1+k_2)^2 B_1 \cosh(\eta_1+\eta_2) + (k_1-k_2)^2 B_2 \cosh(\eta_1-\eta_2)}{B_1 \cosh(\eta_1+\eta_2) + B_2 \cosh(\eta_1-\eta_2)} \right. \nonumber\\ 
	&& \qquad \left. - \left(\frac{(k_1+k_2) B_1 \sinh(\eta_1+\eta_2) + (k_1-k_2) B_2 \sinh(\eta_1-\eta_2)}{B_1 \cosh(\eta_1+\eta_2) + B_2 \cosh(\eta_1-\eta_2)}\right)^2\right] \nonumber\\ 
	&& \qquad + A\left(\frac{(k_1+k_2) B_1 \sinh(\eta_1+\eta_2) + (k_1-k_2) B_2 \sinh(\eta_1-\eta_2)}{B_1 \cosh(\eta_1+\eta_2) + B_2 \cosh(\eta_1-\eta_2)}\right), \\
	Q(x,y,t)&=&2\left[\frac{(k_1+k_2)^2 B_1 \cosh(\eta_1+\eta_2) + (k_1-k_2)^2 B_2 \cosh(\eta_1-\eta_2)}{B_1 \cosh(\eta_1+\eta_2) + B_2 \cosh(\eta_1-\eta_2)} \right. \nonumber\\ 
	&& \qquad \left. - \left(\frac{(k_1+k_2) B_1 \sinh(\eta_1+\eta_2) + (k_1-k_2) B_2 \sinh(\eta_1-\eta_2)}{B_1 \cosh(\eta_1+\eta_2) + B_2 \cosh(\eta_1-\eta_2)}\right)^2\right] \nonumber\\ 
	&& \qquad - A\left(\frac{(k_1+k_2) B_1 \sinh(\eta_1+\eta_2) + (k_1-k_2) B_2 \sinh(\eta_1-\eta_2)}{B_1 \cosh(\eta_1+\eta_2) + B_2 \cosh(\eta_1-\eta_2)}\right).
\eea \label{twosol-ex} \ees 
Note that the above two-soliton solution preserves the combined ($xy$) space-inversion and time-reversal symmetry given by $Q=P(-x+p_0, -y+q_0, -t+r_0)$ as required. 

The above solution (\ref{twosol-ex}) consists of four arbitrary parameters ($k_j$ and $l_j$ with $j=1,2$), along with  three nonlocal real parameters ($p_0,~q_0,~r_0$), and two arbitrary real system parameters ($\alpha$ and $\beta$). Interestingly, the above explicit two-soliton solution involves a combination of both hyperbolic and trigonometric functions for different complex conjugate values of parameters, giving rise to localized and periodic structures during their propagation. 
	
The identities of these two solitons can be controlled by appropriately tuning the available arbitrary parameters, as discussed in the one-soliton case. Hence one can obtain the occurrence and absence of the hyperbolic and trigonometric functions appearing in the solution (\ref{twosol}). In return, we can witness several physically attractive nonlinear wave profiles that result in different types of dynamical characteristics, and we shall discuss each of them in the forthcoming part. 
	
\subsubsection{Elastic Collision of Solitons}
One of the fundamental properties of the two-soliton solution (\ref{twosol}) is to study the collision behaviour. In the present enKP model, the two solitons undergo an elastic collision without changing their identities, such as amplitude, width, and velocity, except for a slight phase shift. Notably, we can categorize the collision scenario into two branches based on the vanishing and non-vanishing background parameter $A$. 
	
\begin{figure}[ht]
{\hfill $P=Q$ \hfill $P$ \hfill \qquad $Q$\hfill \hfill}\\
\centering
\includegraphics[width=0.99\linewidth]{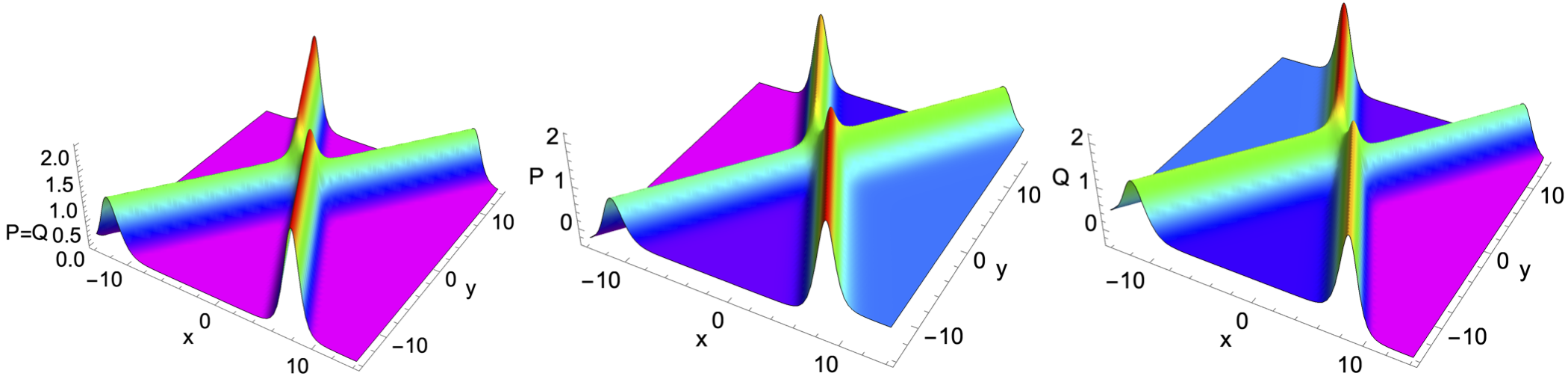}
\caption{Head-on interaction of two solitons in the enKP system (\ref{abekp}) undergoing elastic collisions. Left panel $P=Q$: Elastic collision of symmetric solitons for $A=0$. Middle $P$ and Right $Q$ panels: Elastic collision of bright solitons with induced amplitude by kink--anti-kink background when $A=0.2$. The other parameters are chosen as $\alpha=-0.5$, $\beta=1.0$, $k_1=1.0$, $k_2=0.75$, $l_1=0.5$, $l_2=-0.75$, and $p_0=q_0=r_0=1.0$ at a given time $t=0.5$.}
\label{fig-2soli-1}
\end{figure}

Similar to the one-soliton case, the present two-soliton solution described by (\ref{twosol}) and (\ref{twosol-ex}) admit symmetric and degenerate ($P=Q$) bell-type bright solitons on a vanishing (zero) background when the parameter $A=0$. For this case, one can easily observe their elastic type head-on collision with a clear phase-shift after interaction as depicted in the left panel of Fig. \ref{fig-2soli-1}. On the other hand, when the background parameter is non-zero $A\neq 0$, we obtain localized bell-type solitons appearing on a kink--anti-kink background in $PQ$ components, which arise asymmetrically in their amplitude and localization. Here also, the basic bell-solitons undergo only elastic collision and the kink--anti-kink background induces a small (observable) change in their amplitude as shown in the middle and right panels of Fig. \ref{fig-2soli-1} for $A=0.2$. On increasing the parameter $A$ we can observe a steady increase in the kink and anti-kink background on which the bright solitons exist which consequently modifies the amplitude of asymmetric solitons before and after the interaction. 
	
\subsubsection{Generation of Soliton Bound States}	

\begin{figure}[ht]
\centering
\includegraphics[width=0.99\linewidth]{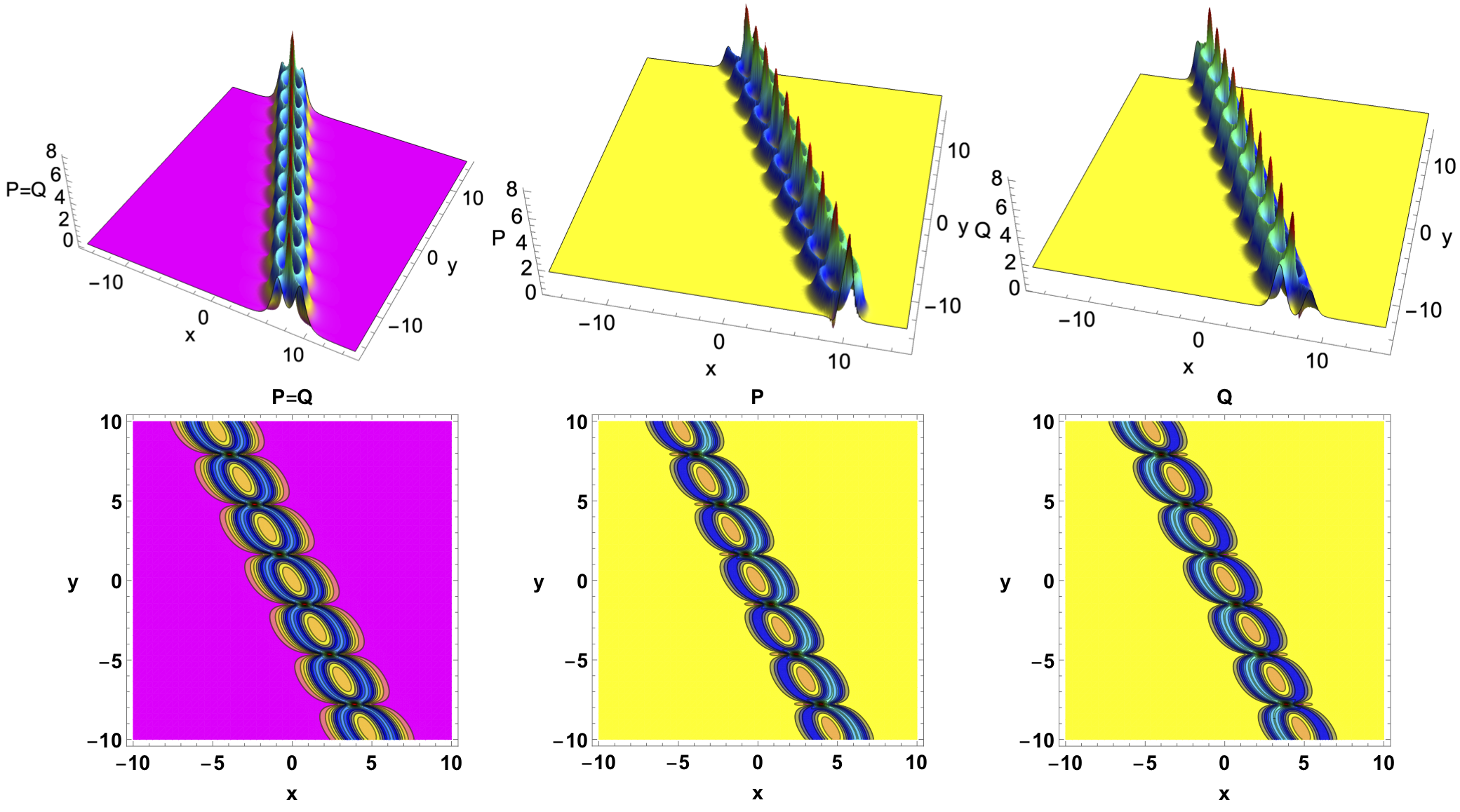}
\caption{Evolution of two soliton bound states resulting in a chain-like pattern in the enKP system (\ref{abekp}). Left panel: Bound state of symmetric profiled solitons for vanishing background $A=0$. Middle \& right panels: Asymmetrically structured soliton bound states for non-vanishing background $A=0.75$. Other parameters are $\alpha=-1.5$, $\beta=1.0$, $k_1=k_2=1.0$, $l_1=1.5+i$, $l_2=1.5-i$, and $p_0=q_0=r_0=0.5$. The bottom panels show the corresponding contour plots. Here and in the forthcoming figures, we have plotted only the absolutes $|P|$ and $|Q|$.}
\label{fig-2soli-bound}
\end{figure}

Beyond the standard elastic collisions discussed above, the two-soliton solution (\ref{twosol-ex}) leads to various other exciting dynamics, which can be achieved through the appropriately chosen soliton parameters. Soliton bound states are a special entity that can be generated when the underlying solitons travel with the same/resonance velocity. Such solitons result in long-lasting interaction without passing through each other and exhibit periodic attraction and repulsion throughout their propagation. The soliton bound states exhibit various features in certain models, including parallel and periodic-type propagation dynamics. The former seems to be both solitons traveling adjacent to each other as there is no contact/interaction. However, for the latter case, one can observe a chain-like pattern formation of the soliton bound-state, which can also be referred to as soliton-chain. In recent years, these resonant velocity solitons have been called soliton molecules too. The present enKP model (\ref{abekp}) admits periodically oscillating and parallel soliton bound states along the spatial coordinates. The dynamics of bound state exhibiting chain structure involving two symmetric and asymmetric solitons are shown in Fig. \ref{fig-2soli-bound} resulting from zero ($A=0$) and non-zero ($A\neq 0$) backgrounds, respectively. One can proceed further to obtain the parallel (bound state) propagation of these symmetric-asymmetric solitons for an appropriate choice of $k_j$ and $l_j$ parameters.   
	
\subsubsection{Formation of Soliton Breathers}	
Apart from the above-explained bound states, the two-soliton solution (\ref{twosol-ex}) exhibits breather characteristics. To understand the dynamics of breathers, let us consider the parametric restriction $k_2=k_1^*$ and $l_2=l_1^*$ (which eventually gives $m_2=m_1^*$). For this choice, one can rewrite the general form of $F$ in (\ref{twosol}a) as below.
\bes \bea 
F_{\text{breather}}(x,y,t) &=& B_1 \cosh\left(E_1(x,y,t)\right)+ B_2  \cos\left(E_2(x,y,t)\right), 
\eea 
where 
$E_1(x,y,t)=2k_{1R}(x - {p_0}/{2}) +2 (k_{1R}l_{1R}-k_{1I}l_{1I}) (y - {q_0}/{2}) + 2 (k_{1R}m_{1R}-k_{1I}m_{1I}) (t - {r_0}/{2})$ and $E_2(x,y,t)=2k_{1I}(x - {p_0}/{2}) +2 (k_{1R}l_{1I}+k_{1I}l_{1R}) (y - {q_0}/{2}) + 2 (k_{1R}m_{1I}+k_{1I}m_{1R}) (t - {r_0}/{2})$. Here $k_{jR}$, $l_{jR}$ and $m_{jR}$ represent the real part of $k_{j}$, $l_{j}$ and $m_{j}$, respectively, while $k_{jI}$, $l_{jI}$ and $m_{jI}$ denote their imaginary parts. 

The above explicit form of $F_\text{breather}$ results in the following two-soliton solution (\ref{twosol}):
\bea
P_{\text{breather}}(x,y,t) &=&\frac{8\left[k_{1R}^2 B_1 \cosh(E_1) - k_{1I}^2 B_2 \cos(E_2)\right]}{B_1 \cosh(E_1) + B_2 \cos(E_2)} - 8\left(\frac{k_{1R} B_1 \sinh(E_1) - k_{1I} B_2 \sin(E_2)}{B_1 \cosh(E_1) + B_2 \cos(E_2)}\right)^2 \nonumber\\ 
	&& \qquad + 2A\left(\frac{k_{1R} B_1 \sinh(E_1) - k_{1I} B_2 \sin(E_2)}{B_1 \cosh(E_1) + B_2 \cos(E_2)}\right),\\
Q_{\text{breather}}(x,y,t) &=&\frac{8\left[k_{1R}^2 B_1 \cosh(E_1) - k_{1I}^2 B_2 \cos(E_2)\right]}{B_1 \cosh(E_1) + B_2 \cos(E_2)} - 8\left(\frac{k_{1R} B_1 \sinh(E_1) - k_{1I} B_2 \sin(E_2)}{B_1 \cosh(E_1) + B_2 \cos(E_2)}\right)^2 \nonumber\\ 
	&& \qquad - 2A\left(\frac{k_{1R} B_1 \sinh(E_1) - k_{1I} B_2 \sin(E_2)}{B_1 \cosh(E_1) + B_2 \cos(E_2)}\right).
\eea \label{breath-ex} \ees 
From the breather solution reduced from the two-soliton soliton, we observed that both participating solitons undergo mutual interaction and give rise to an oscillating wave structure with periodically varying amplitude and position/localization. Remarkably, these breathers can be further classified into the following three types based on the appropriately chosen choices of $k_1$ and $l_1$ parameters:
\begin{itemize}
\item inclined or oblique or general breathers without localization in $x$ or $y$,
\item $x$-localized breathers with periodical variation along $y$, and 
\item $y$-localized breathers with periodical variation along $x$. 
\end{itemize}

\begin{figure}[t]
\centering
\includegraphics[width=0.98\linewidth]{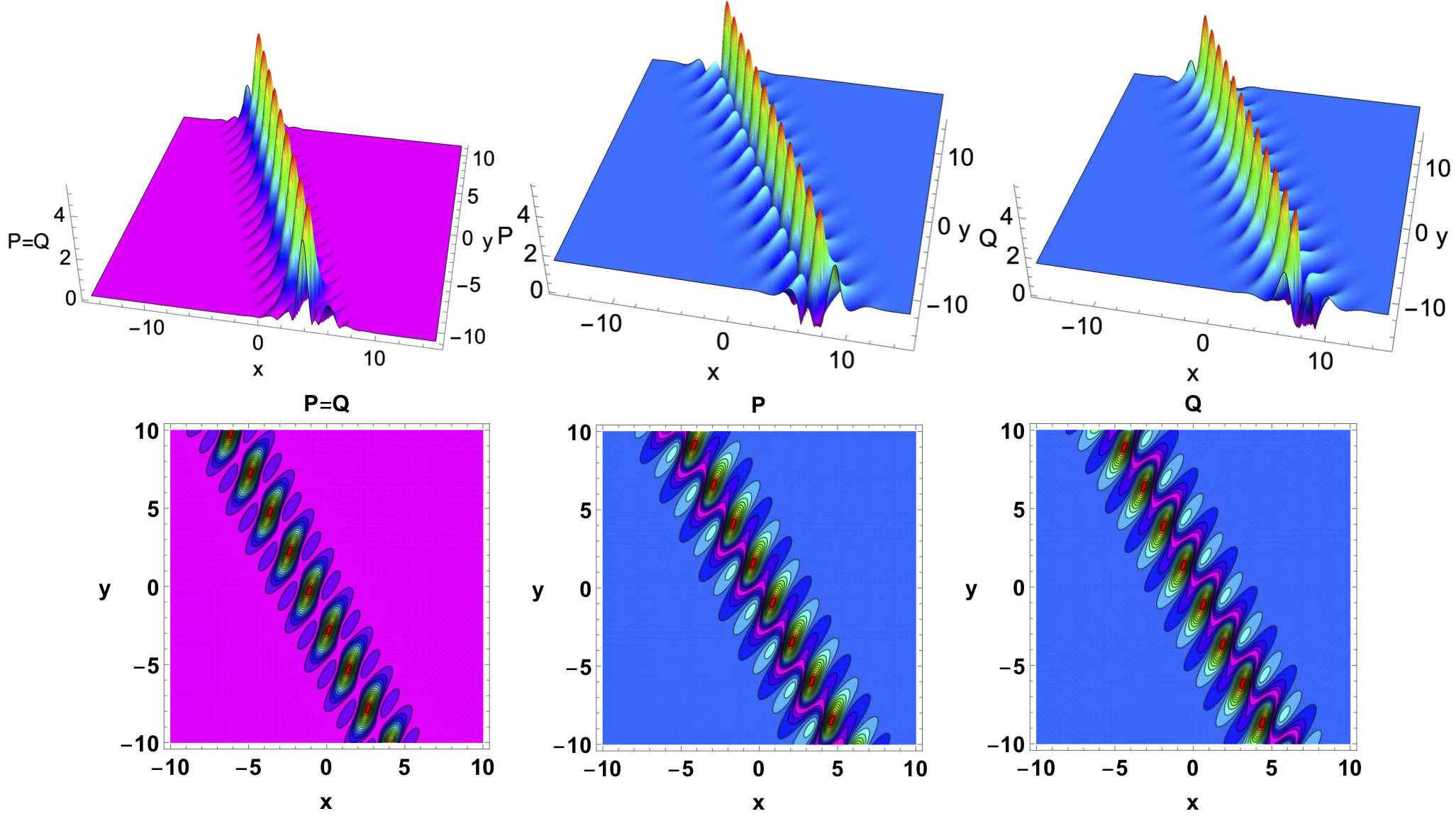}
\caption{Nature of inclined/oblique breathers localized neither in $x$ nor in $y$ for the enKP system (\ref{abekp}) resulting from two-soliton solution (\ref{breath-ex}) for a specific choice of soliton parameters. Left panel: Degenerate type ($P=Q$) symmetric breather on vanishing $A=0$ background. Middle and right panels: Breathers with asymmetric structure ($P\neq Q$) on non-vanishing $A=1.5$ background. Other parameters for both cases are taken as $\alpha=-1.5$, $\beta=1.0$, $k_2=k_1^*=0.5-i$, $l_2=l_1^*=-0.5+0.5i$, and $p_0=q_0=r_0=0.1$ at $t=0.5$. The bottom panels show the corresponding contour plots. } 
\label{fig-inc-breath}
\end{figure}

We can shed more light on these different breathers by analysing the above solution (\ref{breath-ex}). Here the cos-hyperbolic term appearing in Eq. (\ref{breath-ex}a) is responsible for the type of its localization (whether localized in $x$ or $y$), while the cosine function facilitates the periodic oscillation of the breather. To be precise, when $k_{1R}l_{1R}-k_{1I}l_{1I}\neq 0$ and $k_{1R}\neq 0$, we obtain inclined breathers which is neither localized in $x$ nor in $y$. To illustrate the fact more clearly, we have shown such kinds of inclined breathers in Fig. \ref{fig-inc-breath} for particular choices of parameters as given in the captions. 

\begin{figure}[ht]
\centering
\includegraphics[width=0.99\linewidth]{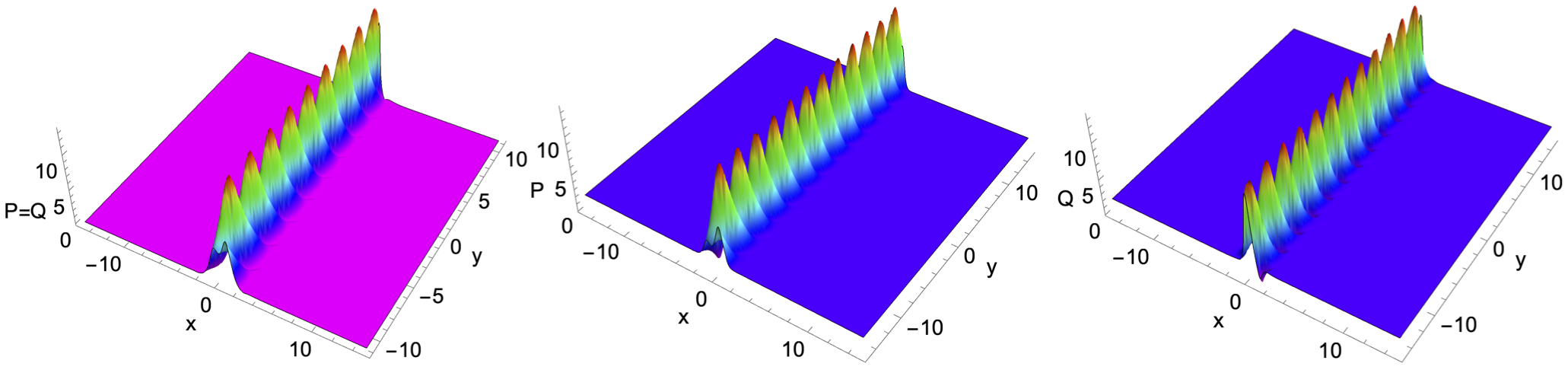}
\caption{Breathers localized in $x$ and periodic in $y$ arising from two-soliton solution (\ref{breath-ex}) for the choice $k_2=k_1^*=1-0.5i$ and $l_2=l_1^*=-0.25+0.5i$. Here we obtain (left panel) degenerate type $P=Q$ symmetric breather on vanishing $A=0$ background and (middle and right panels) asymmetric breathers $P\neq Q$ on non-vanishing $A=1.05$ background with other parameters similar to that of Fig. \ref{fig-inc-breath}.} 		\label{fig-x-breath}
\end{figure}

When $k_{1R}l_{1R}-k_{1I}l_{1I}= 0$ and keep $k_{1R}\neq 0$, the form of $E_1(x,y,t)$ reduces to $E_1(x,t)=2k_{1R}(x - {p_0}/{2}) + 2 (k_{1R}m_{1R}-k_{1I}m_{1I}) (t - {r_0}/{2})$ and the resultant solution becomes simpler than that of Eq. (\ref{breath-ex}b)-(\ref{breath-ex}c). The corresponding solution gives a breather oscillating along $y$ and localized in the $x$ direction, which falls under the second case. Such $x$-localized--$y$-periodic breathers are depicted graphically in Fig. \ref{fig-x-breath} by keeping all other parameters the same as that of Fig. ~\ref{fig-inc-breath}. 

Finally, the third type of breather can be obtained when $k_{1R}= 0$ and keeping $k_{1R}l_{1R}-k_{1I}l_{1I}\neq 0$, the corresponding $E_1(x,y,t)$ form reduces to $E_1(y,t)=-2k_{1I}l_{1I} (y - {q_0}/{2}) - 2 k_{1I}m_{1I} (t - {r_0}/{2})$. The resultant simplified form of $y$-localized breather solution can be written as below from (\ref{breath-ex}b) and (\ref{breath-ex}c).
\bes \bea
P_{y\text{-breather}}(x,y,t) &=&\frac{-2 k_{1I} B_2 \left[4k_{1I} \cos(E_2) + A \sin(E_2)\right]}{B_1 \cosh(E_1) + B_2 \cos(E_2)} - \frac{k_{1I}^2 B_2^2 \sin^2(E_2)}{\left[B_1 \cosh(E_1) + B_2 \cos(E_2)\right]^2}, \quad \\
Q_{y\text{-breather}}(x,y,t) &=&\frac{-2 k_{1I} B_2 \left[4k_{1I} \cos(E_2) - A \sin(E_2)\right]}{B_1 \cosh(E_1) + B_2 \cos(E_2)} - \frac{k_{1I}^2 B_2^2 \sin^2(E_2)}{\left[B_1 \cosh(E_1) + B_2 \cos(E_2)\right]^2}. \quad
\eea \label{y-breath-ex} 
\ees 

\begin{figure}[ht]
\centering
\includegraphics[width=0.99\linewidth]{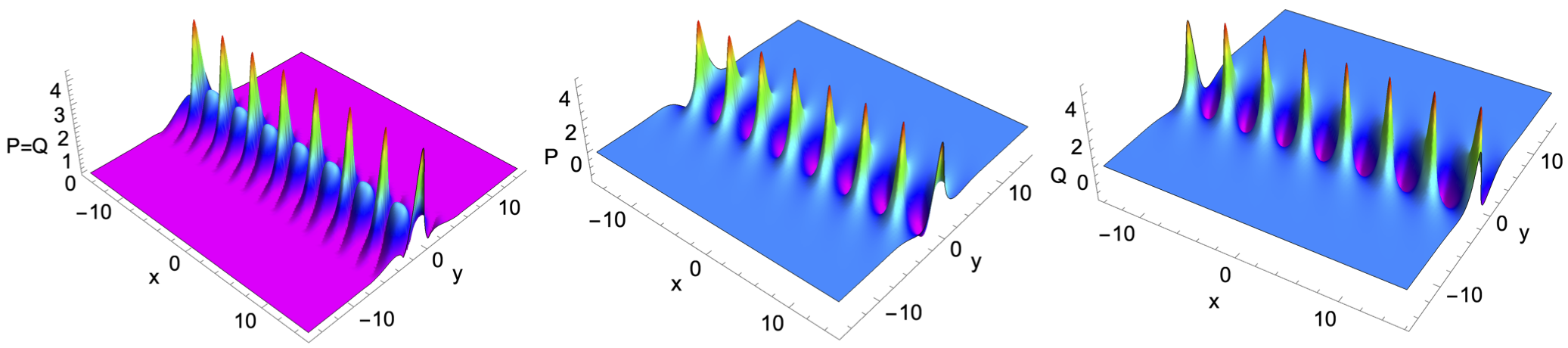}
\caption{The evolution of breathers localized in $y$ and periodic in $x$ at time $t=0.5$ in the enKP system (\ref{abekp}) arising from two-soliton solution for the choice $k_2=k_1^*=-0.75i$ and $l_2=l_1^*=0.5+0.5i$. Here we obtain (left panel) degenerate type $P=Q$ symmetric breather on vanishing $A=0$ background and (middle and right panels) asymmetric breathers $P\neq Q$ on non-vanishing $A=0.75$ background with other parameters similar to that of Fig. \ref{fig-inc-breath}.} 
\label{fig-y-breath}
\end{figure}

\indent 
The above solution represents the breather localized in $y$ and periodic along $x$, which is given in Fig. \ref{fig-y-breath} for completeness. Note that here all three types of breathers admit single-peak structures with symmetric character for vanishing background $A=0$ and asymmetric profiles for non-vanishing background $A\neq 0$. 

Further, one can control other properties of these breathers by suitably tuning the arbitrary parameters by following the above-mentioned relationships. 
	
	
\subsubsection{Dynamics of Periodic and Rational Line Solitons}
	
\begin{figure}[ht]
\centering
\includegraphics[width=0.99\linewidth]{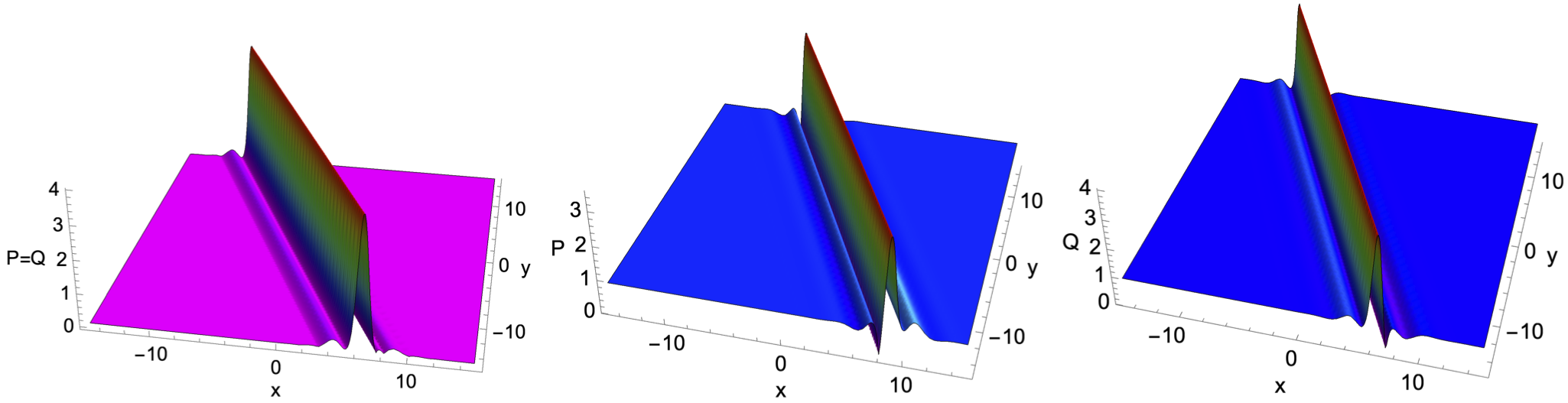}
\caption{The evolution of a rational line soliton arising from the two-soliton solution (\ref{twosol}) of the enKP system (\ref{abekp}). Left panel: Symmetric rational line soliton $P=Q$ for $A=0$. Middle and Right panels: Asymmetric rational solitons $P\neq Q$ for non-vanishing $A=0.75$ background. Other parameters are taken as $\alpha=-1.5$, $\beta=1.0$, $k_1=0.55+i$, $k_2=0.55-i$, $l_1=l_2=0.5$, and $p_0=q_0=r_0=0.1$ at $t=0.5$.}
\label{fig-rat-sol}
\end{figure}

\begin{figure}[ht]
\centering
\includegraphics[width=0.725\linewidth]{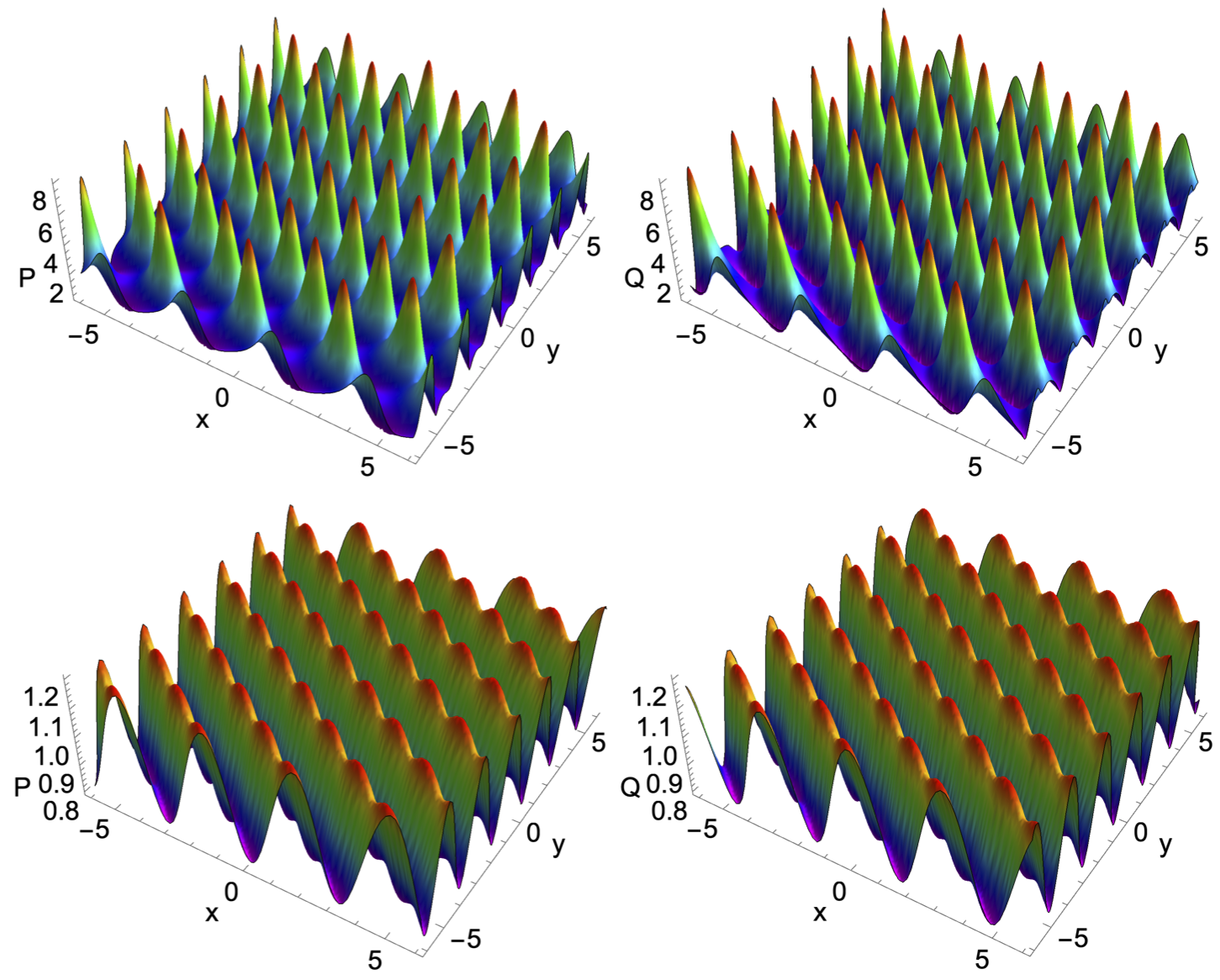}
\caption{The evolution of asymmetric ($P\neq Q$) periodic solitons in the enKP system (\ref{abekp}) arising from two soliton solution for $A=0.5$, $\alpha=-0.5$, $\beta=1.0$, $k_1=i$, $k_2=-i$, $l_1=-1.5$, $l_2=1.5$, and $p_0=q_0=r_0=1.0$ at $t=0$ (top panels) and $t=-0.5$ (bottom panels).} 
\label{fig-perio-sol}
\end{figure}

The final class of nonlinear wave structures we obtain from the two-soliton solution (\ref{twosol}) are rational line solitons and periodic solitons/solutions by suitably tuning the $k_j$ and $l_j$ parameters. These rational line solitons arise when the periodic functions play less or no role; hence, the hyperbolic function gives a localized wave structure. Such symmetric and asymmetric type rational line solitons are shown in  Fig. \ref{fig-rat-sol}. On the other hand, the periodic solitons become significant, with the trigonometric/periodic functions becoming influential in deciding the resulting dynamics, where the hyperbolic functions localising the wave structures vanish and have no effect. For completeness and better understanding, we have depicted asymmetric periodic wave structures of the enKP equation (\ref{abekp}) in Fig. \ref{fig-perio-sol}. 
	
Additionally, one can obtain higher-order (three-, four-, \dots, $N$-) soliton solutions comprising multiple solitons and analyze the phenomena mentioned above, which is beyond the scope of the present work and can be considered for future assignments. Specifically, the study can explain the collision of multiple solitons, the formation of multi-soliton bound states, the interaction between solitons and bound states, and the collision dynamics of breathers with solitons, bound states and breathers. { Upon the construction of $N$-soliton solution, one can investigate the dynamics of different pattern formations leading to T, Y, M, H, and complex web-like wave structures through the resonant mechanism of multiple solitons with long-time interactions. Further, the creation and analysis of other localized waves like breathers, lumps, and rogue waves in the intermediate interaction regime are also of considerable future interest \cite{kspre14, prl2007reso, jpa2022reso, nld2022reso}.}
	
\section{Dynamics of Lump solution}\label{sec-Lump}
	
Lump solutions are localized rational nonlinear waves that generally evolve in higher dimensional systems. This prototype structure gained much attention because of its occurrence in several classes of physical systems. Though the lump can be obtained through the long-wave limit by the reduction of $N$-soliton solution, there are still several nonlinear models for which the construction of lump solutions is tedious because of their non-bilinearizable nature. To overcome this critical issue, Ma has recently proposed a positive quadratic function approach \cite{ zhang2017breather, wuhy2}, which can be successfully applied to both integrable and nonintegrable equations admitting either bilinear or multi-linear form. Even though the present enKP model (\ref{abekp}) admits a compact bilinear equation, we construct a lump solution adopting Ma's approach to test the applicability of the alternate tool \cite{zhang2017breather}. 

In order to find such a lump solution, we choose the following positive quadratic function for $F(x,y,t)$ \cite{wuhy2}:
\bes\bea 
F_{\text{lump}} = \chi_1(x,y,t) ^2 +\chi_2(x,y,t) ^2  + \theta_7, \label{eq13a}
\eea 
where 
\bea 
\chi_1 & =  \theta _1 \left ( x - \dfrac{p_0}{2} \right ) + \theta_2 \left ( y - \dfrac{q_0}{2} \right ) + \theta_3 \left ( t - \dfrac{r_0}{2} \right ),\\
\chi_2 & =  \theta _4 \left ( x - \dfrac{p_0}{2} \right ) + \theta_5 \left ( y - \dfrac{q_0}{2} \right ) + \theta_6 \left ( t - \dfrac{r_0}{2} \right ).
\eea 
Here $\theta_j,~j=1,2,\dots,7,$ are the real parameters to be determined. On substituting $F_{\text{lump}}$ given by (\ref{eq13a}) into the bilinear equation and solving the resulting equation, we obtain 
\bea 
&\theta_1  =& \dfrac{\theta_3 (\theta_2 ^2 - \theta_5 ^2 ) + 2 \theta_2 \theta_5 \theta_6 }{\theta_3 ^2 + \theta_6 ^2 } - \beta \theta_2 - \alpha \theta_3, \\
&\theta_2  =& \dfrac{2 \theta_2 \theta_3 \theta_5  - \theta_2 ^2 \theta_6 + \theta_5^2 \theta_6 }{\theta_3 ^2 + \theta_6 ^2 } - \beta \theta_5 - \alpha \theta_6,\\
&\theta _7=&
\dfrac{3}{(\theta_3^2 + \theta_6^2)(\theta_3 \theta_5 - \theta_2 \theta_6)^2  } 
[\theta_2 ^4 
-2 \beta \theta_2 ^2 \theta_3 + \alpha^2 \theta_3 ^4 + ( \alpha \theta_6 ^2  + \beta \theta_5 \theta_6 - \theta_5 ^2 ) ^2 \nonumber\\ &&+ \theta_3 ^2 (( 2 \alpha + \beta ^2 ) \theta_5 ^2 + 2 \alpha \beta \theta_5 \theta_6 + 2 \alpha ^2 \theta_6 ^2 ) + 2 \theta_2 \theta_3 (\alpha \beta (\theta_3 ^2 + \theta_6 ^2 ) - \beta \theta_5 ^2 - 4 \alpha \theta_5 \theta_6 ) \nonumber\\ &&+ \theta_2 ^2 (( \beta ^2 -2 \alpha ) \theta_3 ^2 + 2 \theta_5 ^2 - 2 \beta \theta_5 \theta_6 + (2 \alpha + \beta )^2 \theta_6 ^2 )]^2,  
\eea \ees 
where the constraint condition $\theta_3 \theta_5 - \theta_2 \theta_6 \neq 0$ has to be satisfied. 

\begin{figure}[ht]
{\hfill P \hfill Q \hfill \qquad P=Q\hfill \hfill}\\
\centering
\includegraphics[width=0.99\linewidth]{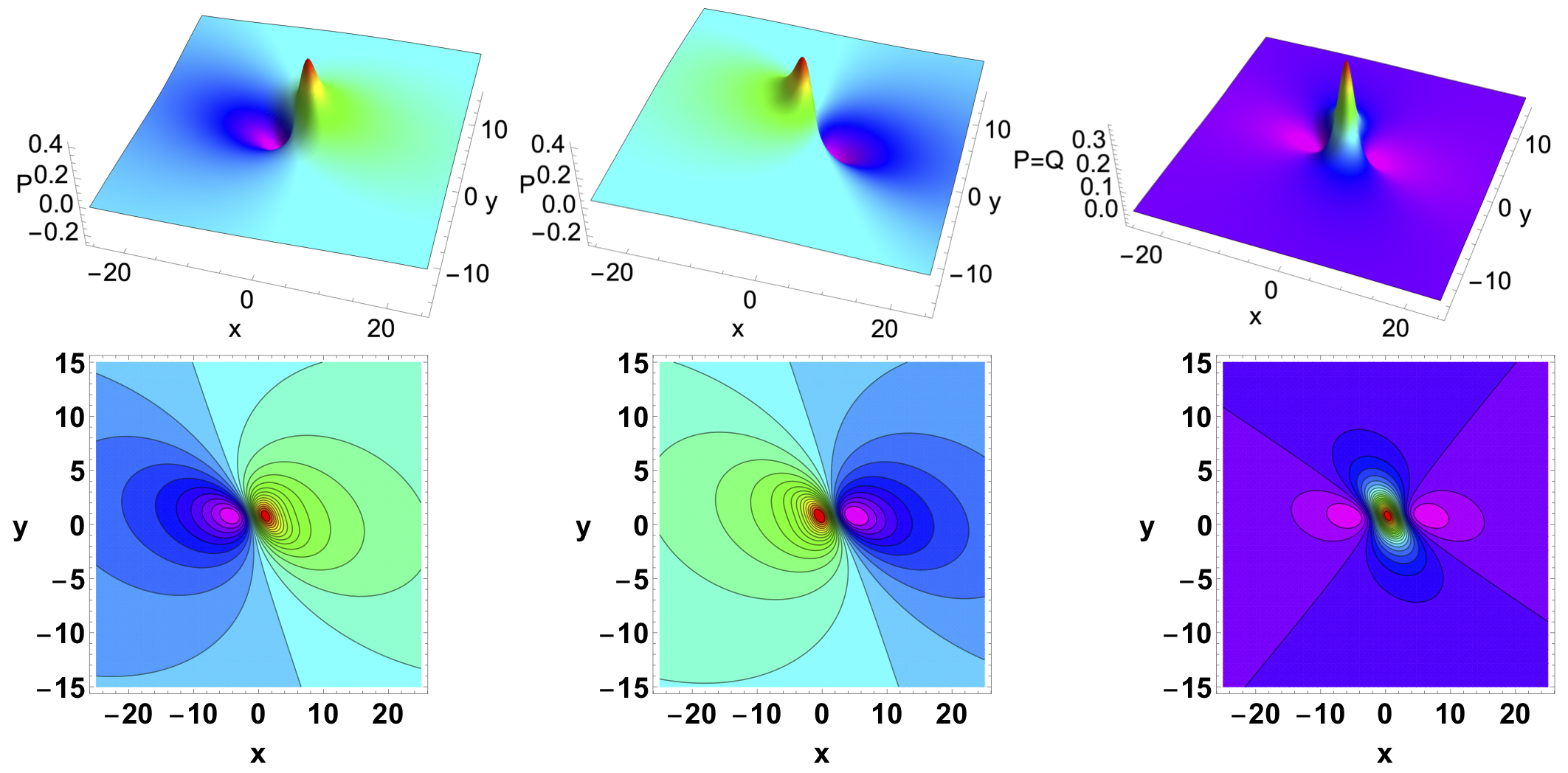}
\caption{Nature of asymmetric (left and middle panels) and symmetric (right panel) single lump solution for the enKP system (\ref{abekp}). The parameters are chosen as $\alpha=0.1$, $\beta=1.5$, $\theta_2=0.5$, $\theta_3=0.5$, $\theta_5=1.5$, $\theta_6=0.75$, and $p_0=q_0=r_0=0.5$ at time $t=0.25$ for (a) asymmetric lump waves in $P$ and $Q$ with $A=0.75$ and (b) symmetric lump wave in $P=Q$ with $A=0$. The bottom panels correspond to their contour plots.}
\label{fig-lump}
\end{figure}

Further, we can obtain the explicit lump form using a variable transformation. The evolutionary dynamics of the obtained lump waves are provided in Fig. \ref{fig-lump}. The received lump waves have four arbitrary parameters with the restriction  $\theta_3 \theta_5 - \theta_2 \theta_6 \neq 0$. It is evident from the demonstration that the lump solution approaches zero when $(x, y)$ approaches infinity. The left and middle panels show asymmetric singly localized bright-dark type lump waves, appearing with the peak bump on the right and left sides, respectively, for non-vanishing background $A\neq 0$. Meanwhile, it is interesting to observe a bright symmetric lump structure shown in the right panel for $A=0$. 
	
Apart from the above first-order lump solution, the occurrence and extraction of higher-order lump solutions to address local and nonlocal characteristics of the present systems involve tedious mathematical derivations, which we have skipped in this work, considering the length of the manuscript. One can study such higher-order lumps and their interaction dynamics with other localized nonlinear waves as open questions and shall form a separate project. 
	
\section{Summary and Conclusions} \label{sec-Conc}
	
To summarize, the study investigates the dynamics of nonlinear waves in a $(2+1)$-dimensional extended nonlocal Kadomtsev-Petviashvili (enKP) model. The enKP model was obtained from the KP equation using the Alice--Bob (AB) approach but was found to be non-integrable according to the Painlev\'e test in its local setting. Using a superposed blinearization technique, we obtained explicit soliton and lump wave solutions that highlight the nonlocal effects on wave structures. The study also explored the wave dynamics of two-soliton solutions, including elastic collisions and breather solutions, and discussed their evolutionary dynamics with graphical demonstrations. 
	
The results of our study demonstrated the impact of nonlocality on the formation of nonlinear waves, resulting in both symmetrical and asymmetrical wave patterns. A single soliton solution highlights the emergence of symmetrical and asymmetrical waves from a combination of bell-type and kink-anti-kink solitons due to the nonlocal background and symmetrical bell-type solitons. Our analysis of the two-soliton solution revealed a range of fascinating wave dynamics, including elastic collisions that can occur as head-on or oblique interactions. By adjusting the soliton parameters ($k_j$ and $l_j$, where $j=1,2$), we discovered soliton bound states undergoing periodic attraction and repulsion without intersecting.
	
We have made a significant discovery in our study by obtaining an explicit form of breather solution from the reduction of the two-soliton solution for specific values of $k_2=k_1^{\ast}$ and $l_2=l_1^{\ast}$. This breather solution encompasses various forms, such as inclined or oblique breathers, localized breathers in both the $x$ and $y$ directions, rational line solitons, and periodic solitons/structures. To provide a complete understanding of the solution, we have briefly analyzed its evolutionary dynamics and presented clear graphical illustrations. Furthermore, we obtained a lump wave solution through the use of a quadratic function as a seed solution and discussed its behavior. The results of the study contribute to the understanding of localized waves in nonlocal nonlinear models and have the potential to be generalized to other integrable and non-integrable nonlocal models.
	
\setstretch{1.20}
\subsection*{Conflict of Interest Statement}
The authors declare that the research was conducted in the absence of any commercial or financial relationships that could be construed as a potential conflict of interest.
	
\subsection*{Author Contributions}
KS: Conceptualization, Formal Analysis, Investigation, Visualization, Writing--Original Draft Preparation, Writing--Review and Editing, Supervision. SS: Conceptualization, Methodology, Writing--Review and  Editing. NK: Resources, Writing--Review and Editing, Funding acquisition. All authors contributed to the discussion of the outcome, manuscript preparation and revision. They read and approved the submitted version. 

\subsection*{Funding}
NK is supported by the National Research Foundation (NRF) of Korea and funded by the Korean Ministry of Science, Information, Communications, and Technology (MSICT) through Grant No.~NRF-2022-R1F1A1-059817 under the scheme of Broadening Opportunities Grants--General Research Program in Basic Science and~Engineering. KS is supported by the Asia-Pacific Center for Theoretical Physics (APCTP), Korea, through the Young Scientist Training (YST) program. 
	
\subsection*{Acknowledgments}
The author K. Sakkaravarthi acknowledges the APCTP, which is supported by the Korean Government through the Science and Technology Promotion Fund and Lottery Fund and the local government of Pohang-si, Gyeongsangbuk-do, Republic of Korea. 

\subsection*{Data Availability Statement}
All required data related to the results are given in the manuscript. 
		
\bibliographystyle{Frontiers-Vancouver}

\end{document}